\documentclass[prb,citeautoscript,amsmath,twocolumn,amssymb,superscriptaddress]{revtex4-1}
\usepackage[utf8]{inputenc}
\usepackage{amsmath}
\usepackage{graphicx}
\usepackage[separate-uncertainty=true,multi-part-units=single]{siunitx}
\usepackage{physics}
\usepackage{mhchem}
\usepackage{xspace}
\usepackage[dvipsnames]{xcolor}
\usepackage[colorlinks, citecolor={blue}, linkcolor={black}]{hyperref}
\AtBeginDocument{\usepackage{booktabs}}

\newcommand{\wte}{\ce{WTe2}\xspace}
\newcommand{\panela}{\textbf{a}\xspace}
\newcommand{\panelb}{\textbf{b}\xspace}
\newcommand{\panelc}{\textbf{c}\xspace}
\newcommand{\paneld}{\textbf{d}\xspace}
\newcommand{\panele}{\textbf{e}\xspace}
\newcommand{\panelf}{\textbf{f}\xspace}

\renewcommand{\figurename}{\textbf{Fig.}} 

\newcommand{\figtitle}[1]{\textbf{#1}\xspace} 
\newcommand{\supplbl}{Ext.~Data~Fig.} 

\begin{document}
\title{ Direct measurement of ferroelectric polarization in a tunable semimetal }

\author{Sergio C. de la Barrera} \email{These authors contributed equally.}
\author{Qingrui Cao} \email{These authors contributed equally.} 
\author{Yang Gao} 
\affiliation{Department of Physics, Carnegie Mellon University, Pittsburgh, PA 15213}
\author{Yuan Gao} 
\affiliation{Department of Physics, Carnegie Mellon University, Pittsburgh, PA 15213}
\affiliation{International Center for Quantum Design of Functional Materials (ICQD), Hefei National Laboratory for Physical Sciences at the Microscale, University of Science and Technology of China, Hefei, Anhui 230026, China}
\author{Vineetha S. Bheemarasetty} 
\affiliation{Department of Physics, Carnegie Mellon University, Pittsburgh, PA 15213}
\author{Jiaqiang Yan} 
\affiliation{Materials Science and Technology Division, Oak Ridge National Laboratory, Oak Ridge, TN 37831, USA}
\author{David G. Mandrus} 
\affiliation{Materials Science and Technology Division, Oak Ridge National Laboratory, Oak Ridge, TN 37831, USA}
\affiliation{Department of Materials Science and Engineering, University of Tennessee, Knoxville, TN 37996, USA}
\affiliation{Department of Physics and Astronomy, University of Tennessee, Knoxville, TN 37996, USA}
\author{Wenguang Zhu} 
\affiliation{International Center for Quantum Design of Functional Materials (ICQD), Hefei National Laboratory for Physical Sciences at the Microscale, University of Science and Technology of China, Hefei, Anhui 230026, China}
\author{Di Xiao} 
\author{Benjamin M. Hunt} \email{bmhunt@andrew.cmu.edu}
\affiliation{Department of Physics, Carnegie Mellon University, Pittsburgh, PA 15213}

\begin{abstract}
Ferroelectricity, the electrostatic counterpart to ferromagnetism, has long been thought to be incompatible with metallicity due to screening of electric dipoles and external electric fields by itinerant charges.
Recent measurements, however, demonstrated signatures of ferroelectric switching in the electrical conductance of bilayers and trilayers of \wte, a semimetallic transition metal dichalcogenide with broken inversion symmetry \cite{fei2018ferroelectric}.
An especially promising aspect of this system is that the density of electrons and holes can be continuously tuned by an external gate voltage.
This degree of freedom enables investigation of the interplay between ferroelectricity and free carriers, a previously unexplored regime.
Here, we employ capacitive sensing in dual-gated mesoscopic devices of bilayer \wte to directly measure the spontaneous polarization in the metallic state and quantify the effect of free carriers on the polarization in the conduction and valence bands, separately.
We compare our results to a low-energy model for the electronic bands and identify the layer-polarized states that contribute to transport and polarization simultaneously.
Bilayer \wte is thus shown to be a canonical example of a ferroelectric metal and an ideal platform for exploring polar ordering, ferroelectric transitions, and applications in the presence of free carriers.
\end{abstract}

\maketitle

Polar materials exhibit charge separation in the absence of an applied electric field, an effect of broken inversion symmetry and a unique polar axis in the crystal \cite{anderson1965symmetry,rabe2007background}.
In certain polar systems, the charge polarization can be switched by an external electric field, an effect known as ferroelectricity.
In principle, the presence or absence of ferroelectric effects depends only on the crystal class and not the details of the electronic structure.
Despite this, nearly all known conventional ferroelectrics are electrically insulating.
Since the first theoretical proposals for ferroelectric metals in 1965 \cite{anderson1965symmetry}, only a handful of experimental claims of ferroelectric-like phases in metallic systems have been reported \cite{shi2013ferroelectric,kim2016polar,benedek2016ferroelectric,rischau2017ferroelectric,yuan2019roomtemperature}, and no clear case for a canonical ferroelectric metal has emerged.
Many such claims fail to demonstrate two key signatures of ferroelectric behavior, direct evidence of the polarization and ferroelectric switching, due to bulk screening effects.

\begin{figure}
    \centering
    \includegraphics[width=3.375in]{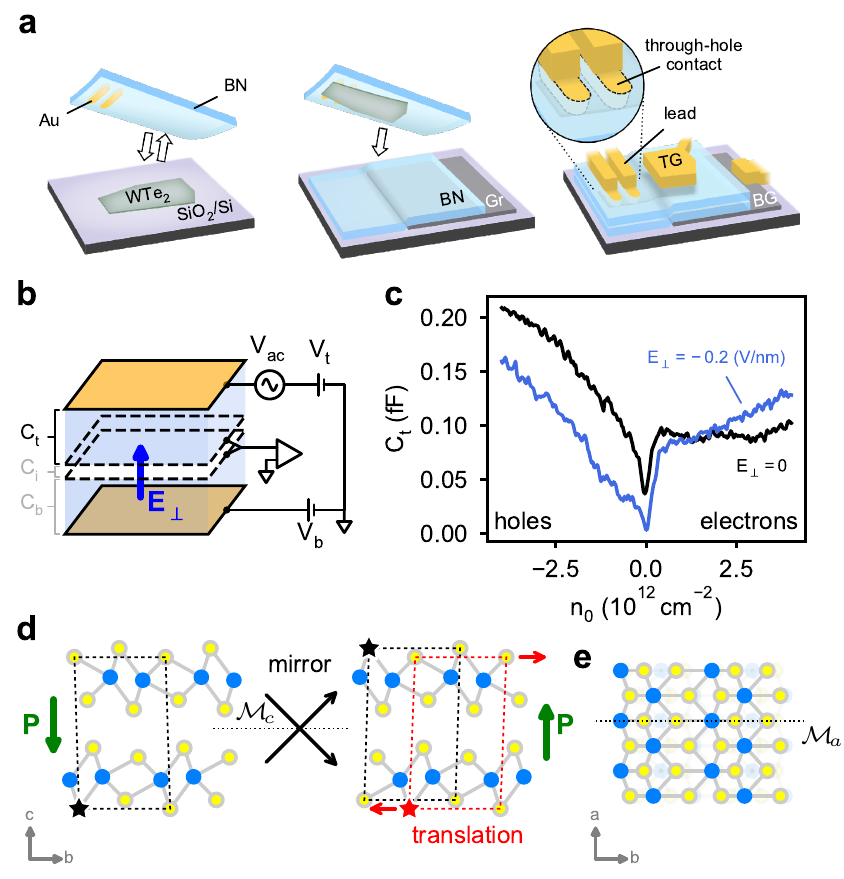}
    \caption{\figtitle{Fabrication and measurement of the bilayer capacitance device.}
    \panela Schematic of our lithography-free encapsulation and contact method, using a boron nitride (BN) crystal previously prepared with through-hole Au contacts to pick up and transfer \wte on to another BN dielectric layer with a graphite bottom gate below. Top gate and leads to the through-hole contacts and bottom gate are patterned after fully encapsulating the \wte.
    \panelb Measurement schematic showing the measurable capacitances: $C_t$, between the top gate and \wte, and $C_b$, between the bottom gate and \wte. $C_i$ is the interlayer capacitance across the bilayer.
    \panelc Measured top capacitance, $C_t$, as a function of carrier density, $n_0$, at zero and finite electric field, $E_\perp$.
    \paneld Side-view structure of bilayer \wte in two stable configurations, each showing the net polarization state along $c$-axis ($\mathbf{P}$).
    The two states are related equivalently by a mirror operation along the $c$-axis ($\mathcal{M}_c$) or by lateral translation between the layers along the $b$-axis of \SI{0.72}{\angstrom} \cite{yang2018origin,liu2019vertical}, or $\SI{\sim 11}{\percent}$ of the unit cell.
    The $\star$ symbol labels a \ce{Te} atom before and after the mirror operation (black) and translation (red) as a visual guide.
    The mirror/translation equivalence allows a subtle shift between the layers to switch the structure between polarization states.
    \panele Top-view structure showing the only invariant symmetry of the crystal, mirror reflection along the $a$-axis ($\mathcal{M}_a$).
    }
    \label{fig:schematic}
\end{figure}

Here, we focus on the polar, semimetallic van der Waals crystal, T$_\text{d}$-\wte, in the limit of two atomic layers, thin enough to admit an external electric field (Fig.~\ref{fig:schematic}\paneld-\panele).
Few-layer crystals of \wte have drawn recent interest for exhibiting a wide variety of low-temperature phases \cite{wu2018observation,fatemi2018electricallytunable,sajadi2018gateinduced,xu2018electricallyswitchable,ma2019observation}.
Recent transport measurements showed that bilayer (2L) and trilayer (3L) \wte exhibit intrinsic, switchable electrical polarization in the conducting state \cite{fei2018ferroelectric}, and separately, surfaces of bulk \wte crystals display hysteresis in piezoresponse force microscopy \cite{sharma2019roomtemperature}.
Subsequent first-principles calculations indicated that the net polarization points only in the out-of-plane direction, and that the underlying mechanism results from a subtle interlayer sliding between the layers in two stable configurations \cite{yang2018origin,liu2019vertical}.
These findings are exciting given the semimetallic and tunable nature of bilayer \wte, which enables reaching both electron and hole bands by electrostatic gating in the ferroelectric state.
While the hysteretic behavior observed in previous experiments is promising, a direct measurement of the metallic polarization is still missing.
Due to methodological limitations it was previously not possible to measure the polarization while varying a pure electric field.
More importantly, these limitations also prevented observing the effect of free carriers on the polarization and its dependence on carrier density, a fundamental open question for ferroelectric metals.

In this work, we directly measure the charge polarization and electronic compressibility as a function of density for electrons and holes with independent control of the electric field.
We study the simplest polar \wte system, a bilayer, via capacitive sensing in a dual-gated, planar capacitance device (Fig.~\ref{fig:schematic}\panelb).
Capacitance measures the electronic compressibility (and thus metallicity) of a 2D system.
In a bilayer 2D system the top-gate and bottom-gate capacitances provide a direct measurement of the \emph{layer-specific} charge distribution \cite{young2011capacitance}, and thus the out-of-plane polarization.
Furthermore, the parallel-plate geometry enables this charge sensing with simultaneous and independent control of the vertical electric field and the carrier density in the bilayer by electrostatic gating.

Our devices each consist of a bilayer \wte crystal encapsulated by two hexagonal boron nitride (hBN) dielectric layers, with metallic top and bottom gates, and contacts integrated into the top hBN layer \cite{telford2018via} (Fig~\ref{fig:schematic}\panela).
We measure the capacitance between the top gate and the bilayer, $C_t$, while applying DC voltages to the top and bottom gates to tune the (nominal) total carrier density, $n_0 \propto C_t^0 V_t + C_b^0 V_b$, and out-of-plane electric field, $E_\perp \propto C_t^0 V_t - C_b^0 V_b$ (Fig.~\ref{fig:schematic}\panelb).
While the geometric contributions to the capacitance ($C_t^0$ and $C_b^0$) are constant, there are additional contributions to the measured capacitance $C_t$ from the electronic compressibility.
In a bilayer system with partial electric field penetration, the layer-specific densities $n_i$ can differ between the two layers ($i=1,2$) for a given total density $n=n_1+n_2$, particularly in the presence of an electric field.
The electrostatic potentials of each layer $\phi_i$ depend on the top and bottom gate voltages, and crucially, any built-in electric field in the bilayer \cite{young2011capacitance,hunt2017direct}. 
As such, $\phi_i$ in each layer can also differ, even in thermodynamic equilibrium.
The electronic compressibility of a 2D bilayer is generally described by a $2\times2$ matrix, $\nu_{ij}=-\partial n_i/\partial\phi_j$, however, in a weakly-coupled bilayer it is possible to characterize the system with only the diagonal elements, $\nu_{ii}$, the layer-specific compressibilities.
Due to the van der Waals nature of the interlayer coupling, in bilayer \wte this is indeed the case (\supplbl~\ref{fig:nu12}). 
Subsequently, the top capacitance may be written,
\begin{equation}
    C_t \approx C_t^0 \left( 1 - \frac{C_t^0}{e^2\nu_{11}} \right), \label{eq:ct-simple}
\end{equation}
(see Methods for details).
The second term in Eq.~\ref{eq:ct-simple} is a quantum correction to the the geometric capacitance $C_t^0$, inversely proportional to the layer-specific compressibility of the top layer, $\nu_{11}$.
The layer-specific densities $n_1$ and $n_2$ can be obtained by integrating the capacitance, and thus the polarization, proportional to $n_1 - n_2$, can be measured.

\begin{figure}
    \centering
    \includegraphics[width=3.375in]{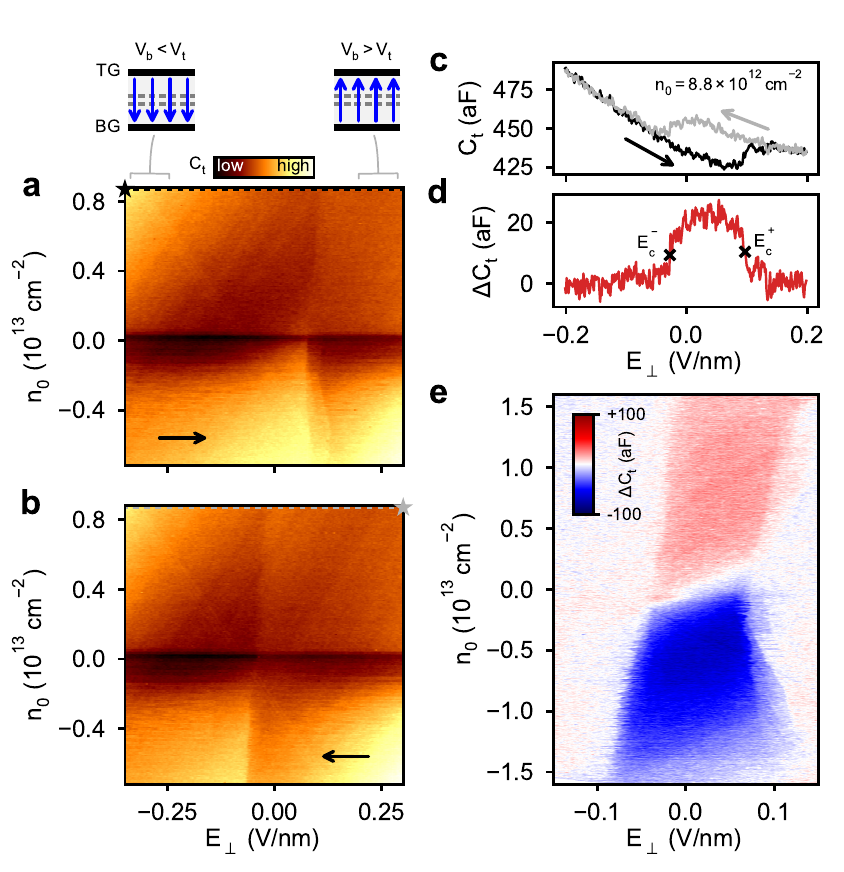}
    \caption{\figtitle{Hysteresis in the electric field response for electrons and holes.}
    \panela Forward and \panelb backward scans of the top capacitance $C_t$ as a function of electric field, $E_\perp$, for a range of carrier densities, $n_0$.
    \panelc Capacitance traces measured along the dashed lines in \panela-\panelb (beginning at the $\star$ symbol in each case) displaying a smooth background as well as sudden jumps at electric field values that depend on the sweep direction.
    \paneld Difference between the traces in \panelc.
    \panele Compilation of differences between forward and backward scans in \panela-\panelb for an extended range of carrier densities, showing the change in sign of the switching from electrons to holes and the gradual density dependence of the switching behavior.
    }
    \label{fig:maps}
\end{figure}

For fixed, external field $E_\perp=0$, the measured $C_t$ as a function of the electron density exhibits a minimum near charge neutrality, $n_0\approx 0$ (Fig.~\ref{fig:schematic}\panelc).
The minimum in $C_t$ indicates the presence of a small band gap (incompressible state) in the \wte bilayer, a feature consistent with previous observations of a sharp drop in the conductance of bilayer \wte in transport \cite{fei2017edge,fei2018ferroelectric}.
At fixed, negative electric field the capacitance minimum becomes more prominent, suggesting that the gap is electric-field-tunable (see \supplbl~\ref{fig:gap}), similar to bilayer graphene \cite{young2012electronic,hunt2017direct}.
However, in contrast to bilayer graphene the electric field response is robustly asymmetric around $E_\perp=0$, as shown in the full $n_0$ and $E_\perp$ dependence in Fig.~\ref{fig:maps}\panela.
This effect results from the absence of an inversion center or mirror plane between the layers of bilayer \wte, implying the crystal is polar along the $c$-axis (Fig.~\ref{fig:schematic}\panele).

To probe the switching behavior of the polar direction, we sweep the electric field back and forth at fixed density, as shown in Fig.~\ref{fig:maps}\panela-\panelb.
Sudden changes are observed in $C_t$ at all densities:
$C_t$ jumps at a positive critical electric field value $E_c^+$ when sweeping toward positive $E_\perp$ (Fig.~\ref{fig:maps}\panela), whereas the critical field $E_c^-$ is negative when sweeping in the negative direction (Fig.~\ref{fig:maps}\panelb), forming a hysteresis loop (Fig.~\ref{fig:maps}\panelc) at each density.
Taking the difference of two representative sweeps in opposite directions $\Delta C_t \equiv C_t^\leftarrow - C_t^\rightarrow$ (Fig.~\ref{fig:maps}\paneld), we see that $C_t$ overlaps nearly everywhere excluding the hysteretic region between the critical fields, $E_c^\pm$, where $C_t$ is multi-valued.
The critical fields generally fall within $|E_\perp| \lesssim \SI{0.1}{V/nm}$ and appear to be weakly dependent on charge density (Fig.~\ref{fig:maps}\panele).
Interestingly, the sign of the hysteretic difference switches for holes ($\Delta C_t < 0$) compared to electrons ($\Delta C_t > 0$), and the magnitude of the difference decreases at large densities of either sign.

\begin{figure*}
    \centering
    \includegraphics[width=6.75in]{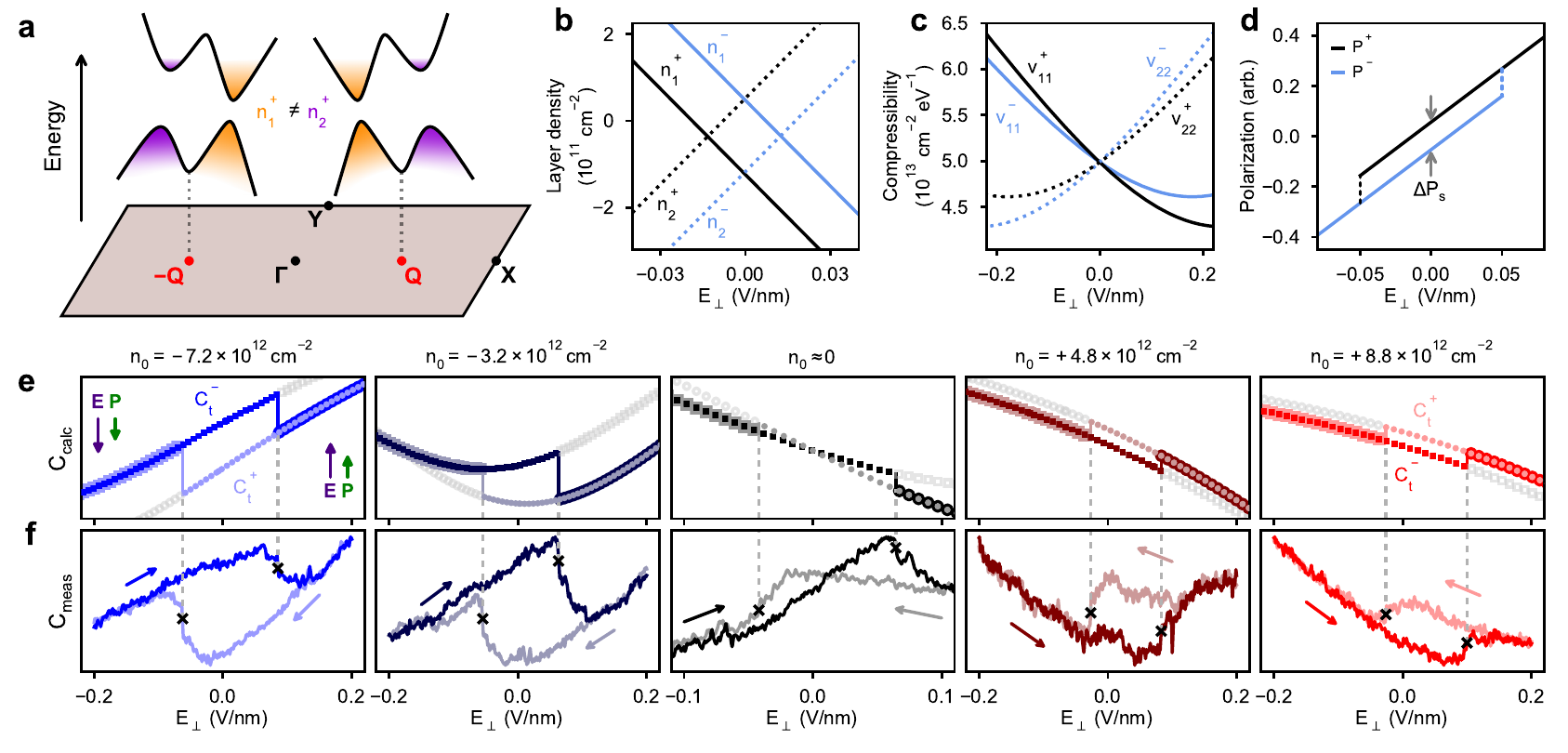}
    \caption{\figtitle{Layer-polarized bands yield distinct capacitance branches for each polarization state.}
    \panela Schematic low-energy bands showing layer-polarized valleys (orange and purple) in the $P^+$ state, for small electron density and $E_\perp=0$. In the $P^-$ state, the colors and layer polarization would be interchanged. Each pair of valleys is centered around a point along $\Gamma$--X, labeled Q. Band parameters are exaggerated to emphasize separation of valleys.
    \panelb Representative calculated layer densities, \panelc compressibilities, and \paneld polarization $\propto n_1^p - n_2^p$ in each polarization state, $p=\pm$, from $n_0\approx 0$.
    \panele Computed top capacitances, $C_t^+$ and $C_t^-$ in the $P^+$ and $P^-$ states, respectively, for the listed densities versus electric field, with colored symbols indicating the hysteretic path observed in experiment, while gray symbols denote portions of each capacitance branch that are inaccessible in experiment due to switching behavior (switching fields are not computed in the model; solid vertical lines are shown extending from experimental switching fields in panel \panelf to reflect the loop observed in experiment). Capacitance is calculated using computed layer densities, potentials, and compressibilities to evaluate Eq.~\ref{eq:ct-full}.
    \panelf Measured top capacitance hysteresis loops for matching electron and hole densities in \panele.
    }
    \label{fig:model}
\end{figure*}

Switching behavior is clearly present in the capacitance, but to understand how this relates to the polarization we must first recognize that the ground state structure of bilayer \wte possesses two stable configurations with opposite polarization \cite{yang2018origin,liu2019vertical} (Fig.~\ref{fig:schematic}\paneld).
Similarly, from Fig.~\ref{fig:maps}\panela-\panelc we see that there are two stable values for $C_t$ in the central region between the two critical fields, $E_c^\pm$.
To reveal how the $C_t$ bistability relates to the two polarization states, we employ a self-consistent calculation of the capacitance using a $\mathbf{k\cdot p}$ model for the bilayer \wte bands \cite{du2018band} together with an exact form of Eq.~\ref{eq:ct-simple} (see Eq.~\ref{eq:ct-full} in Methods).
The $\mathbf{k\cdot p}$ Hamiltonian (Eq.~\ref{eq:hamiltonian}) provides the low-energy electronic structure, shown schematically in Fig.~\ref{fig:model}\panela, and layer-projected wavefunctions in each polarization state.
The layer-specific densities $n_i$ (Fig.~\ref{fig:model}\panelb) and electric potentials $\phi_i$ are obtained from the bands by performing a self-consistent electrostatics calculation for the parallel-plate system (see Methods).
The compressibilities (Fig.~\ref{fig:model}\panelc) are obtained by numerical differentiation and inserted into the general form of Eq.~\ref{eq:ct-simple} to compute the capacitance (given by Eq.~\ref{eq:ct-full}).
Since there are two stable configurations for the bilayer, these quantities all depend on the polarization state $p=\pm$, yielding different values $n_i^p$, $\phi_i^p$, and $\nu_{ii}^p$ in each case.
Using these calculated quantities, the polarization at fixed $n_0$ can be determined from the difference $P^\pm = ed_i ( n_1^\pm - n_2^\pm)$ for interlayer separation $d_i$.
Evaluating this directly, we can construct a schematic polarization loop that illustrates both the dielectric polarization response as well as the spontaneous polarization from the computed layer densities (shown for $n_0\approx0$ in Fig.~\ref{fig:model}\paneld, with switching behavior added manually for illustration).
The difference between the layer imbalance in the two polarization states yields the change in the spontaneous polarization, $\Delta P_s \equiv P^+ - P^-$, as indicated for $E_\perp=0$ in Fig.~\ref{fig:model}\paneld.

\begin{figure}
    \centering
    \includegraphics[width=3.375in]{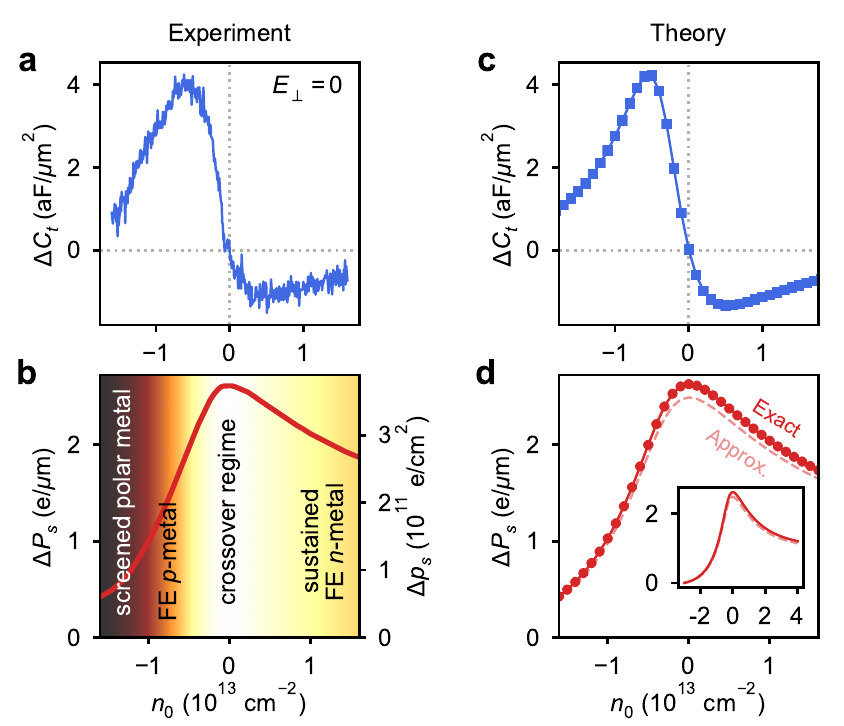}
    \caption{\figtitle{Ferroelectric polarization in the presence of free carriers.}
    \panela Measured density dependence of $\Delta C_t \equiv C_t^+ - C_t^-$ at $E_\perp=0$.
    \panelb Change in measured spontaneous polarization calculated by integration, $\Delta P_s \propto \int \Delta C_t\, dn_0$. Left axis provides 2D polarization units while the right scale is given in terms of charge separation between the layers, $\Delta p_s = \Delta P_s/d_i$ for \wte interlayer separation, $d_i=\SI{0.7}{nm}$. Background shading follows the magnitude of $\Delta P_s$, illustrating distinct regions of ferroelectric (FE) behavior.
    \panelc Computed $\Delta C_t$ and \paneld $\Delta P_s$ based on model Hamiltonian for bilayer \wte, with inset in \paneld showing computed $\Delta P_s$ for an extended density range (with identical units). $\Delta P_s$ is computed by two different methods: ``Exact'' using Eq.~\ref{eq:exactpol}, and ``Approx.'' using Eq.~\ref{eq:approxpol}.
    }
    \label{fig:polarization}
\end{figure}

Whereas polarization switching reflects a reversal of the density imbalance between the layers (Fig.~\ref{fig:model}\panela-\panelb), capacitance switching reflects an inequivalence of the compressibility of the nearest layer in the two polarization states (Fig.~\ref{fig:model}\panelc).
For instance, in Fig.~\ref{fig:model}\panele, we show calculated curves for $C_t$ as a function of $E_\perp$ at a few selected electron and hole densities using the calculated compressibilites.
The result of the calculation is two capacitance curves, $C_t^+$ and $C_t^-$, that correspond to the polarization states $P^+$ and $P^-$, respectively.
Comparing to the experiment (Fig.~\ref{fig:model}\panelf), for sufficiently large positive (negative) $E_\perp$, we only measure the $C_t^+$ ($C_t^-$) branch.
In the hysteretic region, the polarization state $P^+$ or $P^-$ depends on the direction of the electric field sweep, and thus the measured capacitance jumps between the $C_t^+$ and $C_t^-$ branches at the coercive fields, $E_c^\pm$.
While our model does not include all the details of the electronic structure (\supplbl~\ref{fig:DFT}), and thus we do not expect a perfect match with experiment, the general trends of the capacitance with $E_\perp$ and $n_0$ are captured nicely, for instance, the observation that $C_t^+ > C_t^-$ for electrons and $C_t^+ < C_t^-$ for holes.
Next we will see that some of the details of the electronic structure are less relevant for difference quantities such as $\Delta C_t$ and $\Delta P_s$, which leads to improved agreement between the model and experiment, as shown in Fig.~\ref{fig:polarization}.

Finally, we address the density dependence of the polarization to ascertain the effect of adding free charge to a polarized semimetal.
Previously, we defined $\Delta C_t$ as the difference between the right-sweeping and left-sweeping $C_t$ curves.
Having identified the $C_t^\pm$ curves in the central region of each hysteresis loop, we may now equate $\Delta C_t = C_t^+ - C_t^-$ for $E_c^->E_\perp>E_c^+$.
Figure~\ref{fig:polarization}\panela shows the density dependence of $\Delta C_t$ for $E_\perp=0$, equivalent to a vertical line cut from Fig.~\ref{fig:maps}\panele.
The non-monotonic curve follows the size of the $C_t$ hysteresis loop, showing a maximum and minimum on either side of charge neutrality and trending toward zero at large densities.
Due to the switching process acting as a mirror operation on the \wte bilayer, the difference $C_t^+ - C_t^-$ is related by an identity, $\nu_{11}^+(E_\perp) = \nu_{22}^-(-E_\perp)$, to the change in spontaneous polarization (see Methods).
This identity enables extraction of the change in spontaneous polarization for $E_\perp=0$,
\begin{equation}
    \Delta P_s \propto \int \Delta C_t\, dn_0.
\end{equation}

The result of this integration along $E_\perp=0$ is shown in Fig.~\ref{fig:polarization}\panelb.
The spontaneous polarization is strongest at charge neutrality, where there are few free carriers available to screen the built-in polarization.
Adding electrons or holes to the system reveals a pronounced asymmetry in the spontaneous polarization.
Holes appear to screen the polarization much more effectively than electrons for equivalent charge densities.
This final effect can be understood by considering the available states in the conduction and valence bands.
First-principles calculations indicate that the low-energy states are intrinsically layer-polarized (see \supplbl~\ref{fig:DFT}), so despite the fact the the additional charges are free and may conduct current within each layer, they may not simply transfer freely between the layers to screen out the built-in polarization in the absence of an external field.
The wavefunctions of these low-energy states posses strong layer character and thus contribute, in part, to the net polarization rather than fully screening it, contrary to a simple electrostatic picture.
The degree of layer polarization and electronic compressibility differs for electrons and holes in bilayer \wte (as observed in the $e$--$h$ asymmetry of $C_t$ in Fig.~\ref{fig:schematic}\panelc).
Consequently, filling valence band states appears to suppress the net spontaneous polarization whereas filling the conduction band allows polarization to persist to relatively large densities (Fig.~\ref{fig:polarization}\panelb).
Through this mechanism, it is possible to obtain a ferroelectric state that coexists with native metallicity.

In bilayer \wte, the strong layer character of the wavefunctions and asymmetric density of states manifests as several distinct ferroelectric regimes with both bound and free charges contributing to the net polarization.
At large hole densities the measured polarization quickly trends toward zero, entering a screened polar metal regime.
In this phase, there remains underlying polarized bound charge made up of remote states, however, there are enough filled states with opposing layer character to screen this charge, resulting in a suppressed net polarization.
On the other hand, at large electron densities the polarization decreases more slowly, exhibiting a persistent, metallic ferroelectric state.
Near charge neutrality we observe a crossover between these $p$-type and $n$-type metallic ferroelectric states.
The crossover regime is further complicated by the electric field dependence of the conduction and valence bands, as shown in Fig.~\ref{fig:maps}\panela-\panelb.
From $E_\perp=0$, the charge neutral behavior appears to transition from a ferroelectric semimetal with finite band overlap to a ferroelectric insulator with a maximal band gap of $\sim\SI{3}{meV}$ from $\SI{0.3}{V/nm} > |E_\perp| > \SI{0.2}{V/nm}$ (\supplbl~\ref{fig:gap}).

These interpretations are supported by our model calculations, with both the density dependence of calculated $\Delta C_t$ and the integrated $\Delta P_s$ matching well with experiment (Fig~\ref{fig:polarization}\panelc-\paneld).
The model also allows us to investigate the expected behavior over an extended density range, inaccessible in experiment due to dielectric breakdown of hBN.
As shown in the inset of Fig.~\ref{fig:polarization}\paneld, the calculated $\Delta P_s$ continues to exhibit sustained ferroelectric behavior up to electron densities of at least $n_0\approx\SI{4e13}{cm^{-2}}$, whereas the $p$-type ferroelectric behavior is substantially suppressed below $n_0\approx\SI{-2e13}{cm^{-2}}$.

In conclusion, we see that bilayer \wte may be tuned from a sustained ferroelectric $n$-type metal for $n_0 > \SI{e13}{cm^{-2}}$ to a variable-polarization ferroelectric $p$-type metal at intermediate hole densities, followed by a screened polar metal at large hole densities.
In the crossover regime between the $n$- and $p$-type ferroelectric metal states, a ferroelectric insulator or semimetallic phase can be obtained depending on the electric field.
The out-of-plane polarization is largest in the neighborhood of these latter phases, reaching a charge separation of $\sim\SI{3.7e11}{electrons/cm^2}$ over \SI{0.7}{nm} between the layers, equivalent to a volume polarization density of $\sim\SI{0.6}{mC/m^2}$.
The ferroelectric behavior in each metallic phase is supported by the strong layer-polarized character of the states, a result of broken inversion and mirror symmetries combined with weak interlayer coupling.
Together, these observations and the ingredients in \wte that lead to them provide a recipe for engineering and measuring new ferroelectric metal systems.
Looking forward, we anticipate metallic ferroelectricity will manifest in additional transition metal dichalcogenide and van der Waals structures \cite{li2017binary}, enabling a host of new experiments in this previously unexplored regime.

\vfill
\pagebreak

\section{Methods}
\subsection{Fabrication of capacitance devices}

\wte is an especially air-sensitive material, quickly degrading in ambient conditions.
To avoid air exposure during the fabrication process, we first integrate metal contacts into the top \ce{BN} layer and then transfer this template onto the \wte in a \ce{N2}-filled glovebox.
The metal contacts are prepared ahead of time by etching holes through the \ce{BN} crystal and subsequently evaporating pure \ce{Au} contacts to fill these holes.
We pick up the top \ce{BN} along with the integrated \ce{Au} contacts and use this layer to pick up the \wte crystal using standard dry transfer techniques \cite{telford2018via}.
We then pick up a bottom layer of \ce{BN} to fully encapsulate the \wte and place the stack on a graphite or \ce{PdAu} bottom gate, with the \wte partially overlapping the gate but with the through-hole contacts positioned as near as possible to the gated region (Fig.~\ref{fig:schematic}\panela).
At this stage the \wte is sealed from the environment on all sides, allowing us to pattern leads and a top gate using standard electron-beam lithography and metal evaporation.
The top gate is designed to overlap the \wte crystal up to the edge of the bottom gate, aligning carefully to this edge.
This ensures that the entire gated region of the crystal is dual-gated, preventing single-gate features in our capacitance measurements.

\subsection{Capacitance measurement}

Capacitance measurements were performed by connecting the bilayer \wte in each device to a capacitance bridge circuit (\supplbl~\ref{fig:circuit}) with a standard, known capacitor, $C_\text{std}$.
We apply an \SI{11}{kHz} ac excitation $\delta V_t$ with an rms amplitude from \SI{64}{mV} to \SI{140}{mV} on the top gate of the device and a nearly of-out-phase signal $\delta V_\text{std}$ to $C_{std}$ in order to null the total ac signal at the bridge balance point.
Both the amplitude and relative phase of $\delta V_\text{std}$ are tuned to produce a null signal.
As the total capacitance of the device changes, deviations from the null voltage are amplified by a high-electron-mobility transistor (HEMT) mounted within a few millimeters of the sample in the cryostat.
We determine the measured top capacitance $C_t$ from the amplified deviation from the null voltage, $C_t = C_\text{std} (\Delta V_\text{null} / \delta V_\text{std})$.
All measurements were performed in a dilution refrigerator with the sample between \SI{100}{mK} (as in Fig.~\ref{fig:schematic}, Fig.~\ref{fig:maps}\panela-\paneld, and Fig.~\ref{fig:model}) to \SI{30}{K} (Fig.~\ref{fig:maps}\panele), though little change was observed in the capacitance features in this range.

\subsection{$\mathbf{k\cdot p}$ model for bilayer \wte}

To compute the layer compressibilities for bilayer \wte, we adopt a $\mathbf{k\cdot p}$ model for the Hamiltonian describing massive Dirac fermions in two layers with spin-orbit and interlayer coupling.
Beginning with a tilted massive Dirac Hamiltonian \cite{du2018band}, we introduce a layer index, $i=1,2$, and electrostatic potential on each layer, $\phi_i$, to include the effect of a vertical electric field as well as the polarization,
\begin{equation}
    \mathcal{\hat{H}}_{i,s} = \phi_i + t \tilde{k}_{x,i} + v (k_y \sigma_x + \eta \tilde{k}_{x,i} \sigma_y) + (m/2 - \alpha k^2) \sigma_z,
    \label{eq:layerham}
\end{equation}
along with a shifted $k_x$-coordinate, $\tilde{k}_{x,i} \equiv k_x + q_i$ that accounts for the position of each layer-polarized valley (from the $Q$-point shown in Fig.~\ref{fig:model}\panelc, for $\mathbf{Q}=Q\mathbf{\hat{k}_x}$).
The Pauli matrices, $\sigma_{x,y,z}$, act in the orbital pseudospin space, $s=\uparrow,\downarrow$ is the spin degree of freedom, $\eta=\pm 1$ is a chiral index, $t$ tilts the Dirac cones along $k_x$, and $m$ gives rise to the gap at charge neutrality ($\alpha k^2$ ensures convergence of $\mathcal{\hat{H}}_i$ as $k\rightarrow\infty$).
For $E_\perp=0$, the electrostatic potentials take fixed values $\phi_i(E_\perp=0) = \phi_i^0$ (see Table~\ref{tab:kdotp}) to account for the spontaneous polarization.
The spin and layer degrees of freedom are coupled in the $\ket{1,\uparrow}$, $\ket{1,\downarrow}$, $\ket{2,\uparrow}$, $\ket{2,\downarrow}$ basis to obtain the effective Hamiltonian,
\begin{equation}
    \mathcal{\hat{H}}^+ = \begin{pmatrix}
    \mathcal{\hat{H}}_{1\uparrow} & \mathcal{\hat{P}} & 0 & \gamma \\
    \mathcal{\hat{P}} & \mathcal{\hat{H}}_{1\downarrow} & \gamma & 0 \\
    0 & \gamma & \mathcal{\hat{H}}_{2\uparrow} & \mathcal{\hat{P}} \\
    \gamma & 0 & \mathcal{\hat{P}} & \mathcal{\hat{H}}_{2\downarrow} \\
    \end{pmatrix},
    \label{eq:hamiltonian}
\end{equation}
by interlayer coupling, $\gamma$, and spin-orbit coupling,
\begin{equation}
    \mathcal{\hat{P}} = \begin{pmatrix}
    \lambda k_x - i \lambda_y k_y & 0 \\
    0 & -\lambda_x k_x - i \lambda_y k_y \\
    \end{pmatrix},
    \label{eq:soc}
\end{equation}
where $\lambda_x$, $\lambda_y$ govern the spin-orbit coupling strength in the $x$ and $y$ directions.
Such a model has previously been applied to describe the Berry curvature dipole observed in bilayer \wte in vertical electric fields \cite{ma2019observation} and captures the spin and shifted-valley character of \wte bands determined by first-principles methods \cite{du2018band,muechler2016topological}.
Here, we explicitly include a polarization state label in the Hamiltonian, $\mathcal{\hat{H}}^p$, $p=\pm$, to denote the two possible configurations for the \wte bilayer.
The Hamiltonian for the opposite polarization state, $\mathcal{\hat{H}}^-$, is obtained by interchanging layers, $1\leftrightarrow 2$, and spin.
Note that both the electrostatic potential, $\phi_i$, and shifted valley coordinate, $\tilde{k}_{x,i}$, depend on the layer index and thus are interchanged upon a simultaneous mirror operation ($\mathcal{M}_c$) and inversion of the electric field.
As a result, the Hamiltonian possesses an identity,
\begin{equation}
    \mathcal{\hat{H}}^+(E_\perp) = \mathcal{\hat{H}}^-(-E_\perp),
    \label{eq:hamidentity}
\end{equation}
relevant for extracting the polarization from the measured capacitance, to be discussed in a subsequent section.

\subsection{Capacitance calculation}

To understand the various features of the measured capacitance, we apply this effective model to calculate the band structure and eigenstates and numerically evaluate the capacitance for a similar geometry.
Starting with Eq.~\ref{eq:hamiltonian}, we determine the probability densities in each layer and sum over occupied states to obtain the layer-specific densities, $n_1$ and $n_2$ for the top and bottom layers, respectively.
The layer-specific compressibility elements $\nu_{ij}$ are determined by taking partial derivatives of the densities $n_i$ with respect to the electric potentials on each layer.
The resulting compressibilities in conjunction with the geometric capacitances then determine the calculated capacitances.

The eigenstates of Eq.~\ref{eq:hamiltonian} span an $8\times 8$ Hilbert space of layer, spin, and orbitals.
We obtain layer densities by first calculating the probability density on each layer from the eigenstates,
\begin{equation}
    {\vert \psi_i(\mathbf{k}) \vert}^2 = \sum_{s,m} \braket{ism}{ism},
\end{equation}
for spin $s=\uparrow,\downarrow$ and orbital pseudospin $m$.
Note that here, and for the remainder of this section we have dropped explicit labels making reference to the polarization state for simplicity.
Layer densities are then determined by integration,
\begin{equation}
    n^{\prime}_i = \int \frac{d^2 k}{(2 \pi)^2} f(\mathbf{k}) {\vert \psi_i(\mathbf{k}) \vert}^2,
    \label{eq:totalden}
\end{equation}
where $f(\mathbf{p})$ is the Fermi-Dirac distribution.
However, Eq.~\ref{eq:totalden} only includes contributions from valence electrons. The contribution from ions can be determined as follows. 
With the absence of an external electric field, $E_\perp=0$ and $\phi_i=\phi_i^0$, and the bilayer is at charge neutrality.
We then have a vanishing total charge, $n_1^\prime+n_2^\prime+n_1^{ion}+n_2^{ion}=0$.
On the other hand, the two layers have same ionic composition.
As a result,
\begin{equation}
    n_1^\text{ion} = n_2^\text{ion} = -\frac{1}{2} \left[ n^{\prime}_1(\phi_1^0, \phi_2^0) + n^{\prime}_2(\phi_1^0, \phi_2^0) \right].
\end{equation}
From this definition, we obtain the relevant layer densities,
\begin{equation}
    n_i (\phi_1, \phi_2) = n^{\prime}_i (\phi_1, \phi_2) + n_i^\text{ion}.
\end{equation}
To perform a self-consistent calculation, we treat the electrostatic potentials, $\phi_i$, as parameters and compute the layer densities, $n_i = n_i(\phi_1,\phi_2)$ for an appropriate grid of values.
The compressibilities are then obtained by taking the partial derivatives $\nu_{ij} = \partial n_i/\partial \mu_j = -\partial n_i/\partial \phi_j$ over the same parameter space (where $\phi_i = -\mu_i$ for layer-specific chemical potential $\mu_i$ and with the sample connected to ground).
Finally, the capacitances are defined in terms of small signal variations, $C_t \equiv (\delta n_1+\delta n_2) / \delta V_t$, $C_b \equiv (\delta n_1+\delta n_2) / \delta V_b$, and $C_p \equiv \delta n_t / \delta V_b$, where $C_p$ is the penetration field capacitance, measured from the top gate to the bottom gate, for charge on the top gate $n_t$.
These expressions may be evaluated in terms of the compressibility elements (following Ref.~\citenum{young2011capacitance}),
\begin{subequations}
\label{eq:capacitances}
\begin{align}
    C_t &= C_t^0 \left[1 - \frac{\text{det}(\hat{C})-C_b^0e^2\nu_{21}+C_t^0e^2\nu_{22}}{\text{det}(e^2\hat{\nu}+\hat{C})}\right], \\
    \label{eq:ct-full}
    C_b &= C_b^0 \left[1 - \frac{\text{det}(\hat{C})-C_t^0e^2\nu_{12}+C_b^0e^2\nu_{11}}{\text{det}(e^2\hat{\nu}+\hat{C})}\right], \\
    &\quad C_p = C_b^0 C_t^0 \frac{C_i - e^2\nu_{12}}{\text{det}(e^2\hat{\nu}+\hat{C})},
\end{align}
\end{subequations}
with geometric capacitances $C_t^0$, $C_b^0$, and interlayer capacitance $C_i$, as shown in Fig.~\ref{fig:schematic}\panelb, each given by $C_j = \epsilon\epsilon_0A/d_j$ using $\epsilon=3.2$ for hBN and $\epsilon=1$ within the bilayer.
The compressibility matrix, $\hat{\nu}$, is a symmetric matrix, $\nu_{ij}=\nu_{ji}$, and takes the form
\begin{equation}
    \hat{\nu} = \begin{pmatrix}
    \nu_{11} & \nu_{12} \\
    \nu_{21} & \nu_{22} \\
    \end{pmatrix},
    \label{eq:compmat}
\end{equation}
while $\hat{C}$ is a matrix of geometric capacitances,
\begin{equation}
    \hat{C} = \begin{pmatrix}
    C_i + C_t^0 & -C_i \\
    -C_i & C_i + C_b^0
    \end{pmatrix}.
\end{equation}

To obtain results in terms of $E_\perp$ and $n_0$, we first apply a transformation derived from the charge balance equations,
\begin{subequations}
\begin{align}
    en_1 &= C_t^0 (V_t + \phi_1/e) + C_i (\phi_1 - \phi_2)/e \\
    en_2 &= C_b^0 (V_b + \phi_2/e) + C_i (\phi_2 - \phi_1)/e
\end{align}
\end{subequations}
to determine $V_b$ and $V_t$,
\begin{equation}
    \begin{pmatrix} C_t^0 V_t \\ C_b^0 V_b \end{pmatrix}
    = e\begin{pmatrix} n_1 \\ n_2 \end{pmatrix}
    - \hat{C} \begin{pmatrix} \phi_1/e \\ \phi_2/e \end{pmatrix},
\end{equation}
and then calculate $E_\perp=(V_t/d_t - V_b/d_b)/2$ for top and bottom hBN thicknesses, $d_t$ and $d_b$, respectively, and $n_0=(C_t^0 V_t + C_b^0 V_b)/e$.
Finally, to include the effect of the two polarization states, as shown in the main text for $C_t^\pm$, we compute the capacitances $C_t^p$ separately for each case, $p=\pm$, using the compressibilities, $\nu_{ij}^p$, calculated from the eigenstates of $\mathcal{\hat{H}}^p$, derived from the appropriate form of Eq.~\ref{eq:hamiltonian}.

\subsection{Simplifications and approximations}
In practice, quantum capacitances of 2D electron gases tend to dominate over geometric capacitances from external electrodes and gates due to the small energy spacing between levels.
For instance, the density of states (and thus the non-interacting electronic compressibility) in monolayer graphene is of order $\rho_\text{MLG}(0) \sim \SI{e13}{eV^{-1} cm^{-2}}$, yielding a quantum capacitance of order $C_q \sim \SI{20}{fF/\micro\meter^2}$.
In comparison, a metallic gate separated by \SI{10}{nm} of hBN from a parallel plate generates a geometric capacitance one order of magnitude smaller, $C_g \sim \SI{2}{fF/\micro\meter^2}$.
Typical thicknesses for hBN gate dielectrics are even larger, such as the device shown in the main text with $d_t \approx \SI{15}{nm}$ and $d_b \approx \SI{19}{nm}$, resulting in even smaller geometric capacitances.
Additionally, in each layer of bilayer \wte, screening from carriers in the nearby layer and weak interlayer coupling significantly reduces the off-diagonal compressibilities relative to the diagonal terms in Eq.~\ref{eq:compmat}.
Together, these considerations imply a hierarchy, $|\nu_{11}|, |\nu_{22}| \gg C_t^0, C_b^0, C_i \gg |\nu_{12}|, |\nu_{21}|$ (ignoring factors of $e^2$ preceding the $\nu_{ij}$ terms).
As a result, the diagonal compressibilities $\nu_{ii}$ dominate the behavior of $C_t$ and $C_b$, enabling an approximate form for Eqs.~\ref{eq:capacitances},
\begin{subequations}
\label{eq:capapprox}
\begin{align}
    C_t &\approx C_t^0 \left(1 - \frac{C_t^0}{e^2\nu_{11}}\right), \\
    C_b &\approx C_b^0 \left(1 - \frac{C_b^0}{e^2\nu_{22}}\right),
    \label{eq:cb-simple}
\end{align}
\end{subequations}
as shown in Eq.~\ref{eq:ct-simple} in the main text.
While the numerical computations shown in the paper make use of the full Eq.~\ref{eq:ct-full}, deviations of the approximate formulas from the exact quantities are negligible and thus the simpler form is given in main text Eq.~\ref{eq:ct-simple} to facilitate understanding.

\subsection{Capacitance relation to polarization}
To extract the spontaneous polarization from the measured capacitance, we seek the change in the polarization,
\begin{equation}
    \Delta P_s = P^+ - P^-,
\end{equation}
obtained by integrating the difference of polar response functions,
\begin{equation}
    \Delta P_s = d_i \int \left( \frac{\partial p^+}{\partial n_0} -  \frac{\partial p^-}{\partial n_0} \right) dn_0 + \Delta P_0,
    \label{eq:exactpol}
\end{equation}
defined up to a constant of integration, $\Delta P_0$, which we take to be zero at large hole densities, where the density dependent spontaneous polarization appears to saturate.
The polar response in each state, $\pm$, is defined by
\begin{equation}
    \frac{\partial p^\pm}{\partial n_0} = \frac{\partial n_1^\pm}{\partial n_0} - \frac{\partial n_2^\pm}{\partial n_0}.
    \label{eq:dpdn}
\end{equation}
Eq.~\ref{eq:exactpol} together with Eq.~\ref{eq:dpdn} is employed to compute the ``Exact'' spontaneous polarization in the model calculations shown in Fig.~\ref{fig:polarization}\paneld.
In the experiment, we do not have direct access to the partial derivatives in Eq.~\ref{eq:dpdn}, however the symmetry of our device allows an approximate form of Eq.~\ref{eq:exactpol} to be related to $\Delta P_s$.

Following Ref.~\citenum{hunt2017direct}, we express the polar response in Eq.~\ref{eq:dpdn} as,
\begin{equation}
    \frac{\partial p}{\partial n_0} = \frac{\partial n_1}{\partial V_t} \frac{\partial V_t}{\partial n_0} + 
                                      \frac{\partial n_1}{\partial V_b} \frac{\partial V_b}{\partial n_0} - 
                                      \frac{\partial n_2}{\partial V_t} \frac{\partial V_t}{\partial n_0} - 
                                      \frac{\partial n_2}{\partial V_b} \frac{\partial V_b}{\partial n_0},
\end{equation}
where we have dropped the label indexing the polarization state.
We then employ
\begin{equation}
    e\begin{pmatrix} \delta n_1 \\ \delta n_2 \end{pmatrix}
    = [1 - \hat{C}(e^2\hat{\nu} + \hat{C})^{-1}]
    \begin{pmatrix} C_t^0 \delta V_t \\ C_b^0 \delta V_b \end{pmatrix},
\end{equation}
to obtain the partials $\partial n_i/\partial V_{t,b}$, and finally rewrite the polar response in terms of the average geometric capacitance, $\langle C\rangle = (C_b^0+C_t^0)/2$, and the asymmetry, $\delta = (C_b^0-C_t^0)/(C_b^0+C_t^0)$,
\begin{equation}
\begin{aligned}
    \frac{\partial p}{\partial n_0} &= \frac{C_i}{\langle C\rangle^2} \left(1 + \frac{\langle C\rangle}{2C_i}\right)(C_b - C_t) \\
    +& \frac{4C_i}{\langle C\rangle^2} \left[ \frac{2\langle C\rangle - C_b - C_t}{4} - \left(1 + \frac{\langle C\rangle}{2C_i}\right) C_p \right] \delta + \mathcal{O}(\delta^2).
\end{aligned}
\end{equation}
For the device shown in the main text (Device~A in Table~\ref{tab:devices}), $\delta \approx 0.11$ and $\langle C\rangle/2C_i \approx 0.067$.
Keeping only the leading order terms in these two small parameters, we obtain
\begin{equation}
    \frac{\partial p}{\partial n_0} \approx  -\frac{C_i}{\langle C\rangle^2} (C_t - C_b).
\end{equation}
Therefore, we have
\begin{align}
    \frac{\partial \Delta P_s}{\partial n_0} &= d_i \left( \frac{\partial p^+}{\partial n_0} - \frac{\partial p^-}{\partial n_0} \right) \nonumber \\
    &\approx -d_i \frac{C_i}{\langle C\rangle^2} \left[(C_t^+ - C_b^+) - (C_t^- - C_b^-)\right] \nonumber \\
    &= -d_i \frac{C_i}{\langle C\rangle^2} (\Delta C_t - \Delta C_b),
    \label{eq:dpdnapprox}
\end{align}
where $\Delta C_{t(b)} \equiv C_{t(b)}^+ - C_{t(b)}^-$.
To evaluate this further, we first rearrange Eq.~\ref{eq:ct-simple} and explicitly show the $E_\perp$ dependence (ignoring factors of $e^2$),
\begin{equation}
    \frac{C_t(E_\perp)}{\left(C_t^0\right)^2} \approx \frac{1}{C_t^0} - \frac{1}{\nu_{11}(E_\perp)},
\end{equation}
which implies
\begin{equation}
    \frac{\Delta C_t (E_\perp)}{\left(C_t^0\right)^2} = \frac{C_t^+ (E_\perp)}{\left(C_t^0\right)^2} - \frac{C_t^- (E_\perp)}{\left(C_t^0\right)^2} = \frac{1}{\nu_{11}^- (E_\perp)} - \frac{1}{\nu_{11}^+ (E_\perp)}.
    \label{eq:deltaCt}
\end{equation}
Similarly, we have
\begin{equation}
    \frac{\Delta C_b (E_\perp)}{\left(C_b^0\right)^2} = \frac{1}{\nu_{22}^- (E_\perp)} - \frac{1}{\nu_{22}^+ (E_\perp)}.
    \label{eq:deltaCb}
\end{equation}
Using the fact that switching the polarization state is equivalent to a mirror operation between the layers, we find
\begin{equation}
    \nu_{22}^\pm(E_\perp) = \nu_{11}^\mp(-E_\perp),
    \label{eq:nuidentity}
\end{equation}
a symmetry that is evident in the identity relating the Hamiltonian in the two polarization states, Eq.~\ref{eq:hamidentity}, and the subsequent compressibilities derived from each.
Physically, this implies that measuring the capacitance from one side of the device in the two polarization states is related to a measurement of the capacitance from both sides of the device by geometrical factors.
Thus, we invoke Eq.~\ref{eq:nuidentity} to relate Eq.~\ref{eq:deltaCt} and Eq.~\ref{eq:deltaCb},
\begin{equation}
    \frac{\Delta C_b (E_\perp)}{\left(C_b^0\right)^2} = \frac{1}{\nu_{11}^+ (-E_\perp)} - \frac{1}{\nu_{11}^- (-E_\perp)} = - \frac{\Delta C_t (-E_\perp)}{\left(C_t^0\right)^2},
\end{equation}
or equivalently
\begin{equation}
    \Delta C_b (E_\perp) = -\frac{\left(C_b^0\right)^2}{\left(C_t^0\right)^2} \Delta C_t (-E_\perp) = -\left(\frac{d_t}{d_b}\right)^2 \Delta C_t (-E_\perp),
\end{equation}
for hBN dielectrics on both sides of the device.
Finally, we insert this identity into Eq.~\ref{eq:dpdnapprox},
\begin{equation}
    \frac{\partial \Delta P_s (E_\perp)}{\partial n_0} = -d_i \frac{C_i}{\langle C\rangle^2} \left[\Delta C_t (E_\perp) +\left(\frac{d_t}{d_b}\right)^2 \Delta C_t (-E_\perp)\right],
\end{equation}
solely in terms of measured $\Delta C_t(E_\perp)$.
Integrating this expression from the largest hole density measured, $n_l$, up to $n_0$, we obtain
\begin{align}
    \Delta P_s(E_\perp,n_0) &\approx -d_i \frac{C_i}{\langle C\rangle^2} \int_{n_l}^{n_0} \! \bigg[\Delta C_t (E_\perp, n_0^{\prime}) \nonumber \\
    &\quad + \left(\frac{d_t}{d_b}\right)^2 \Delta C_t (-E_\perp, n_0^{\prime})\bigg] \, \mathrm{d}n_0^{\prime}.
\end{align}
Evaluating at $E_\perp=0$, we arrive at the approximate integral used to determine the measured polarization in Fig.~\ref{fig:polarization}\panelb,
\begin{equation}
    \Delta P_s(0,n_0) \approx -d_i \frac{C_i}{\langle C\rangle^2} \left[1+\left(\frac{d_t}{d_b}\right)^2\right] \int_{n_l}^{n_0} \! \Delta C_t (0, n_0^{\prime}) \, \mathrm{d}n_0^{\prime}.
    \label{eq:approxpol}
\end{equation}
The same expression is employed in Fig.~\ref{fig:polarization}\paneld to calculate the ``Approx.'' curve based on computed capacitance data, illustrating the small deviation from the exact polarization introduced by making the approximations outlined in this section.
A small constant of integration is included in the total measured spontaneous polarization shown in Fig.~\ref{fig:polarization}\panelb, $P_0=\SI{0.42}{e/cm}$, inferred by matching the magnitude of the measured $\Delta P_s(n_0)$ with the calculated curve.
The latter curve exhibits saturation of $\Delta P_s$ at large hole densities (taken to be zero) and thus offers a lower bound on the constant of integration and ultimately the maximum value of the polarization near charge neutrality.

\subsection{First-principles calculations}

As the essential ingredient to interpret the capacitance in experiments, the low-energy model in Eq.~\ref{eq:hamiltonian} is well supported by first-principles calculations.
The most essential element for the capacitance is the polarization or the occupation difference between the top and bottom layer.
In \supplbl~\ref{fig:DFT}, we show such layer character in the band structure near one group of valleys from the first-principles calculations compared with those from the low-energy model.
By tracing the valence band, we find that bands from the low-energy model and first-principles calculation show similar tilting, together with similar layer-occupancy variation.
This striking similarity demonstrates that the low-energy model is indeed capable of capturing the essential physics for the capacitance calculation.

The supporting density functional theory calculations shown in \supplbl~\ref{fig:DFT}\panelb are performed with the Vienna Ab initio Simulation Package (VASP) using the PBEsol functional.
The plane wave basis cutoff is \SI{300}{eV}.
The atomic structures are relaxed until forces on every atom are smaller than \SI{0.01}{eV/\angstrom} and the vacuum layer is larger than \SI{15}{\angstrom}.
The van der Waals (vdW) functional used here is the zero damping DFT-D3 method with small modifications: the vdW correction is only applied to atomic pairs from different layers, since monolayers can be calculated well without vdW correction.
This modification ensures that the structure of the sub-layer of the bilayer will not be affected by an inappropriate vdW correction, leading to the same lattice constant for the monolayer and bilayer system.

\section{Acknowledgements}
We thank Qiong Ma, Kenji Yasuda, Felix L{\"u}pke, Dacen Waters, and Evan Telford for fruitful discussions.
B.M.H., S.C.B., and Q.C.\ were supported by the Department of Energy under the Early Career award program (DE-SC0018115) for all aspects of this project.
Y.G.\ and D.X.\ are supported by the Department of Energy, Basic Energy Sciences, Grant No.~DE-SC0012509.
W.Z. acknowledges support from the National Key Research and Development Program of China (2019YFA0210004).

\section{Author contributions}
S.C.B.\ and B.M.H.\ conceived of the project.
S.C.B.\ and Q.C.\ fabricated the devices and performed the measurements, with fabrication support from V.S.B.
Y.G.\ and Y.G.\ performed the theoretical calculations under the supervision of W.Z. and D.X.
J.Y.\ and D.G.M.\ grew the bulk \wte crystals.
S.C.B., Q.C., and B.M.H.\ analyzed and interpreted the data and wrote the manuscript with contributions from all authors.

\section{Competing interests}
The authors declare no competing interests.

\bibliography{refs}

\clearpage
\setcounter{figure}{0}
\renewcommand{\figurename}{\textbf{Extended Data Fig.}}

\begin{figure*}
    \centering
    \includegraphics[width=3.375in]{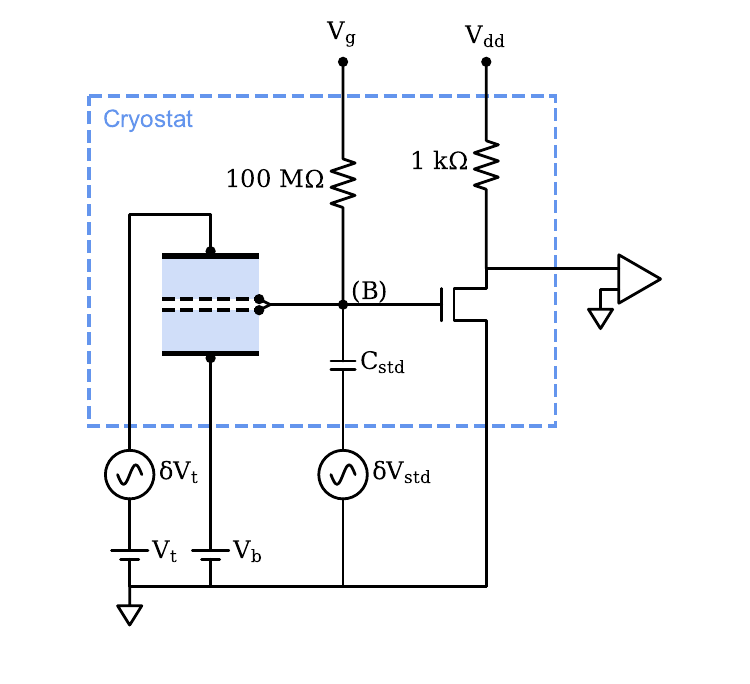}
    \caption{\figtitle{Diagram of the capacitance bridge circuit.}
    The capacitance bridge and amplifier circuit is mounted next to the device in the cryostat sample space (dashed box).
    The circuit is supplied with external dc gate voltages $V_t$, $V_b$ and ac excitation voltages applied to the top gate $\delta V_t$ and the standard capacitor $\delta V_{std}$.
    The deviation from null signal at the bridge balance point (B) is amplified by a high-electron-mobility transistor (HEMT) held at optimal gain by a dc gate voltage $V_g$ and drain current set by $V_{dd}$.}
    \label{fig:circuit}
\end{figure*}

\begin{figure*}
    \centering
    \includegraphics[width=3.375in]{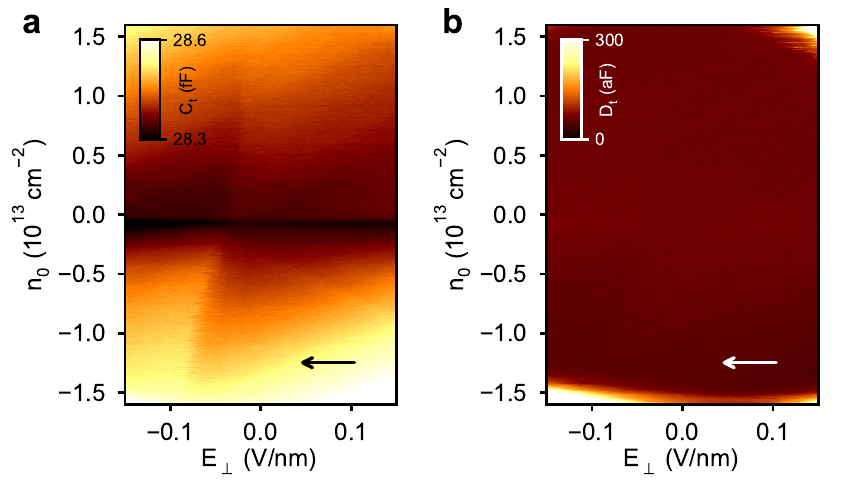}
    \caption{\figtitle{Capacitance and dissipation in the compressibility regime.}
    \panela Capacitance (in-phase) data from a single sweep direction (see arrow) from the same measurement shown in Fig.~\ref{fig:maps}.
    \panelb Dissipation (out-of-phase) signal from the same measurement, showing no features in the relevant parameter space.
    Here the dissipation is offset by \SI{1.23}{fF} to shift the average value of the dissipation to zero.
    The scale of \panelb is set to match \panela for comparison.
    The origin of the shifted mean value of the dissipation is systematic phase shifts in the combined capacitance bridge and cryostat wiring.
    The bright features in the corners of the scan range in the dissipation signal arise from deviations in the cryogenic amplifier operating point, and do not affect the analyses described in the main text.
    }
    \label{fig:loss}
\end{figure*}

\begin{figure*}
    \centering
    \includegraphics[width=6.75in]{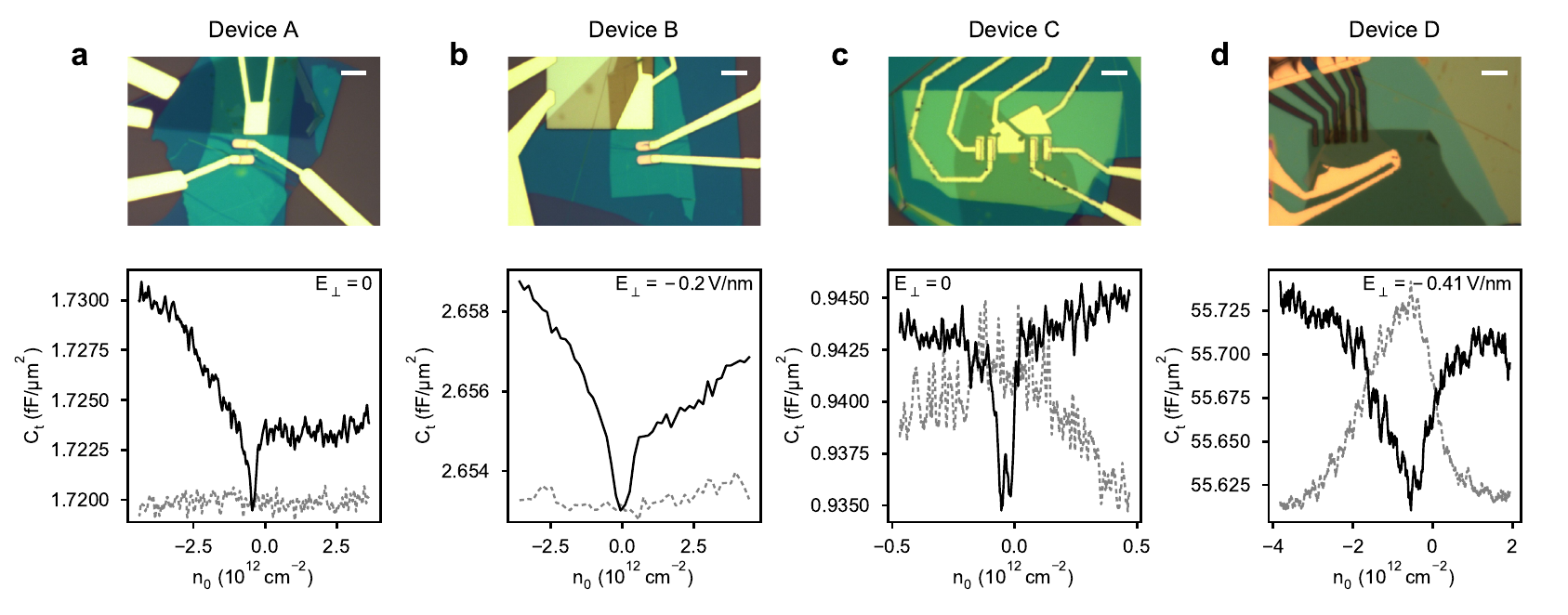}
    \caption{\figtitle{Additional devices in the study.}
    \panela-\paneld Optical images (top) and measured capacitance at fixed electric field for Devices A-D, respectively.
    Gray dashed curves show the dissipation signal, vertically shifted up to the minimum value of the capacitance for comparison.
    All scale bars are \SI{5}{\micro m}.
    }
    \label{fig:four-devices}
\end{figure*}

\begin{figure*}
    \centering
    \includegraphics[width=6.75in]{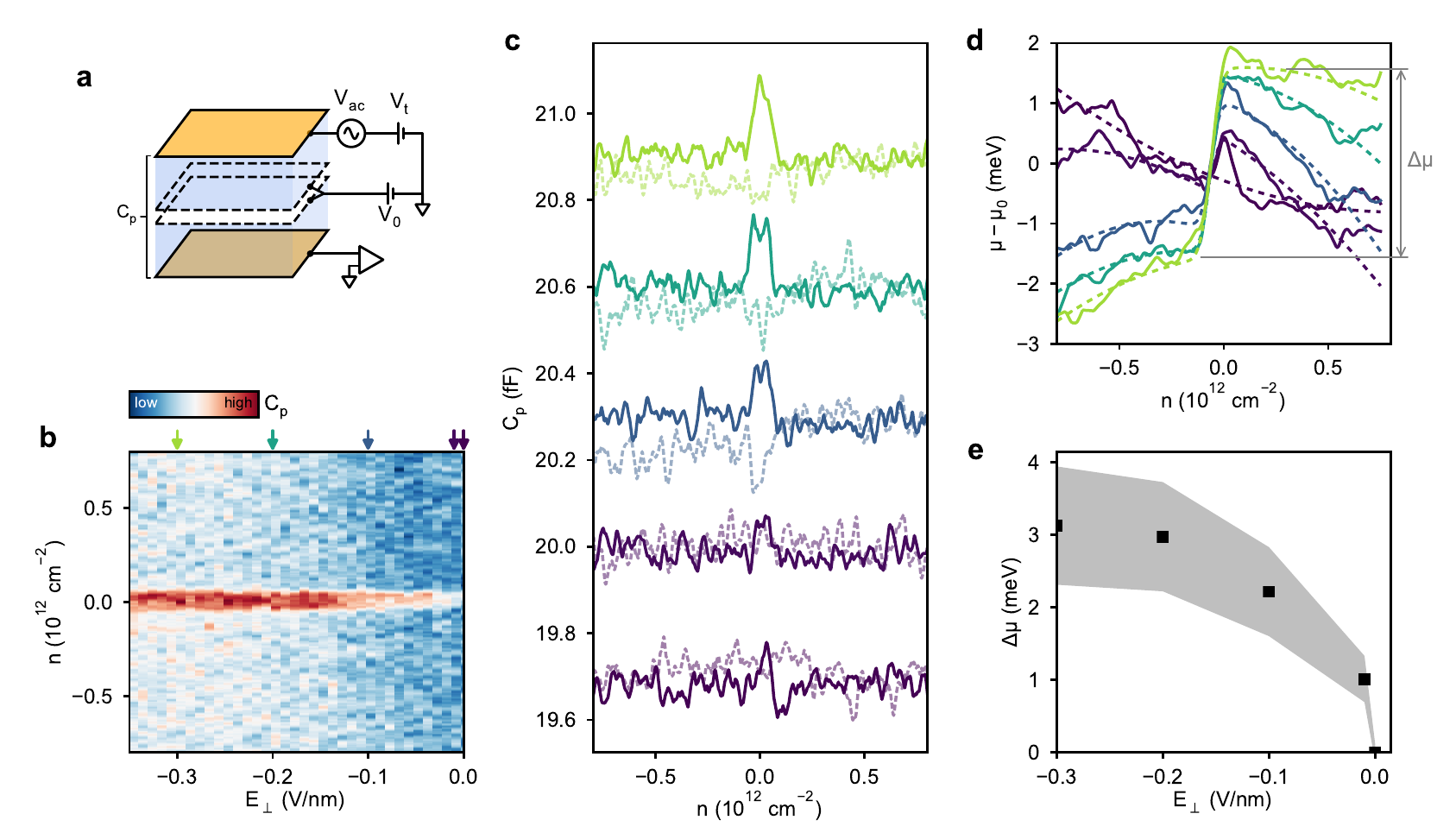}
    \caption{\figtitle{Determination of the electric-field-induced gap.}
    \panela Schematic of the penetration field capacitance $C_p$ circuit, measuring the capacitance between the top and bottom gate.
    \panelb Measured $C_p$ dependence on electric field $E_\perp$ and density $n$, showing an incompressible (large $C_p$) feature at charge neutrality (measured at \SI{4}{K}).
    \panelc Selected $C_p$ traces from \panelb as a function of density at fixed $E_\perp$ (indicated by colored arrows at the top of \panelb), with dashed lines showing the dissipation (out-of-phase) signal for each trace on the same scale. Curves shifted vertically for clarity.
    \paneld Shift of the chemical potential $\mu-\mu_0 = (e/\bar{C}^2) \int C_p\, dn$ calculated by integrating the traces in \panelc with respect to density, using average geometric capacitance $\bar{C}=(C_t^0+C_b^0)/2$ and setting the constant of integration $\mu_0$ to the mid-gap value of the chemical potential. Dashed lines indicate error-function fits used to extract the jump the in chemical potential $\Delta\mu$ at charge neutrality.
    \panele Jump in the chemical potential at charge neutrality extracted from fits in \paneld, a quantitative measure of the electric-induced gap as a function of $E_\perp$. Shaded region indicates the total uncertainty of the extracted values including fitting errors as well as uncertainty in the value of the reference capacitor, $C_\text{std} = \SI{13\pm3}{pF}$.
    }
    \label{fig:gap}
\end{figure*}

\begin{figure*}
    \centering
    \includegraphics[width=3.375in]{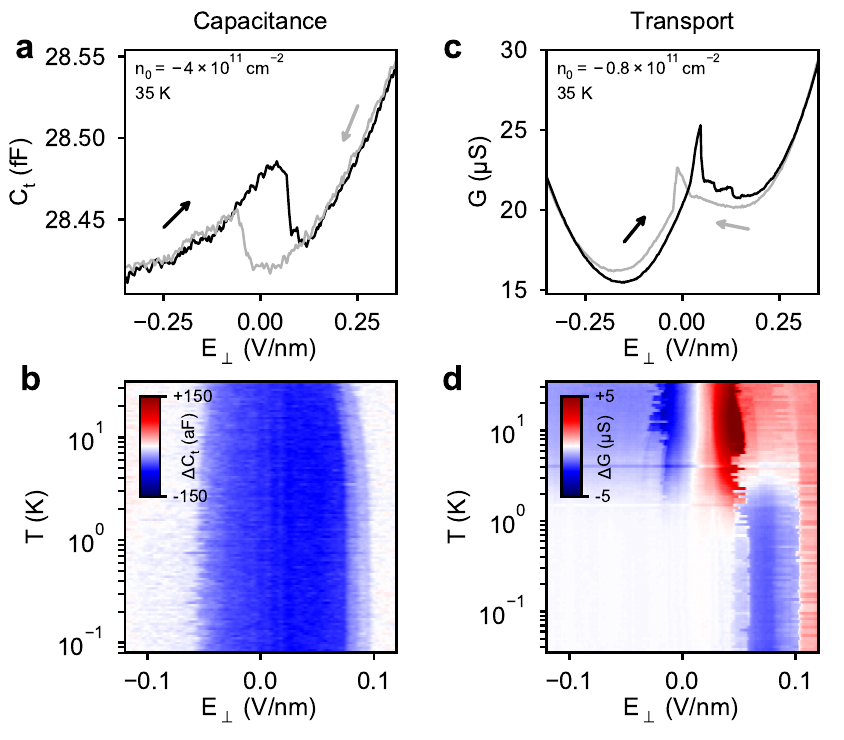}
    \caption{\figtitle{Temperature dependence of the hysteresis.}
    \panela Capacitance hysteresis loop from Device~A at the highest temperature measured, \SI{35}{K}.
    \panelb Temperature dependence of the hysteretic difference $\Delta C_t$ (described in the main text), showing little change in the capacitance signal and critical fields, $E_c^\pm$.
    \panelc Two-terminal conductance from Device~C showing switching behavior at a comparable low hole density to \panela.
    \paneld Temperature dependence of the difference of conductance sweeps, $\Delta G$, showing a weakening signature below \SI{\sim 1}{K}.
    In contrast, capacitance measurements suggest that the polarization is not strongly affected by temperature in this range.
    }
    \label{fig:temperature}
\end{figure*}

\begin{figure*}
    \centering
    \includegraphics[width=3.375in]{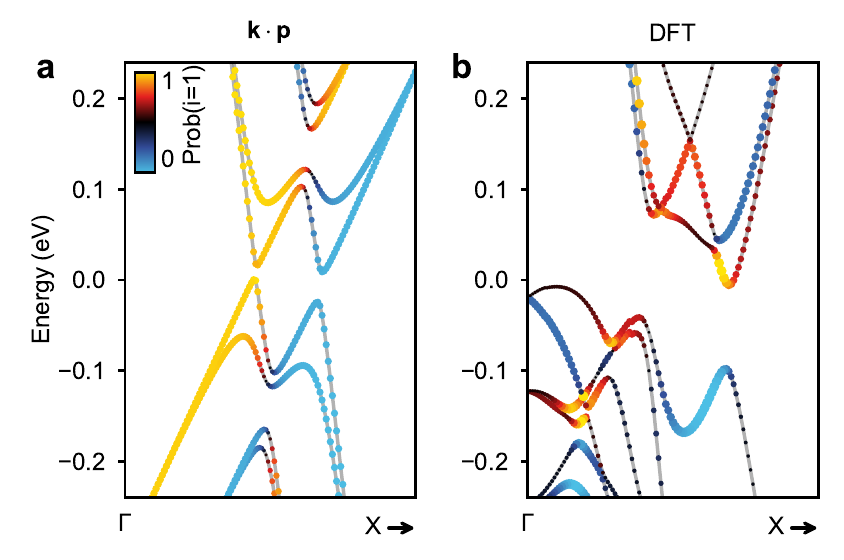}
    \caption{\figtitle{Calculated bands with layer polarization.}
    \panela Calculated bands from the low-energy $\mathbf{k\cdot p}$ model (for $E_\perp=0$) and \panelb density functional theory (DFT).
    Color and symbol size for each eigenstate denotes the normalized probability of appearing in layer 1, $\mathop{\text{Prob}}(i=1)$, according to the color scale in \panela.
    The probability of appearing in layer 2 is complementary, $\mathop{\text{Prob}}(i=2) = 1 - \mathop{\text{Prob}}(i=1)$.
    }
    \label{fig:DFT}
\end{figure*}

\begin{figure*}
    \centering
    \includegraphics[width=3.375in]{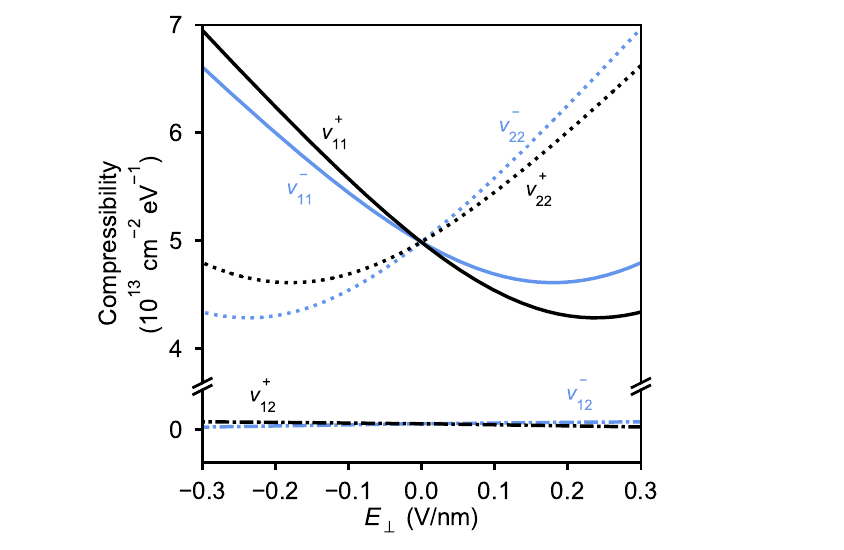}
    \caption{\textbf{Complete compressibility matrix for the bilayer system.}
    Comparison of compressibility magnitudes highlighting the secondary role of differential interlayer couplings $\nu_{12} = \nu_{21} = dn_1/d\phi_2 = dn_2/d\phi_1$.}
    \label{fig:nu12}
\end{figure*}

\begin{table*}[]
    \centering
    \begin{tabular}{ >{\centering}m{20mm} >{\centering}m{20mm} >{\centering}m{20mm} >{\centering}m{20mm} >{\centering}m{20mm} >{\centering\arraybackslash}m{20mm} }
        \toprule 
        Device & $A$~(\si{\micro\meter^2}) & $d_b$~(\si{nm}) & $d_t$~(\si{nm}) & Bottom gate & Top gate \\
        \midrule 
        A & 16.3 & 19 & 15 & Gr & Au \\ 
        B & 40.6 & 41 & 13 & Pt & Au \\ 
        C & 23.7 & 32 & 40 & Gr & Au \\ 
        D & 22.0 & 20 & 30 & Gr & Au \\ 
        \bottomrule
    \end{tabular}
    \caption{Table of device parameters, including area ($A$), bottom and top gate dielectric thicknesses ($d_b$ and $d_t$), and bottom and top gate materials (Gr = graphite). All data shown in the main text were obtained from Device A.}
    \label{tab:devices}
\end{table*}

\begin{table*}[]
    \centering
    \begin{tabular}{ >{}m{30mm} >{\arraybackslash}m{30mm} }
        \toprule 
        $\phi_1^0 = \SI{-0.08}{eV}$ & $\phi_2^0 = \SI{0.03}{eV}$ \\
        $q_1 = 0.1\pi$ & $q_2 = 0.15\pi$ \\
        $\lambda_x = 0$ & $\lambda_y = \SI{0.2535}{eV.\angstrom}$ \\
        $t = \SI{2.5355}{eV.\angstrom}$ & $m = \SI{0.1}{eV}$ \\
        $\gamma = \SI{0.05}{eV}$ & $\eta_1 = -\eta_2 = -1$ \\
        $\alpha = \SI{2.857}{eV.\angstrom^2}$ & \\
        \bottomrule
    \end{tabular}
    \caption{Parameters employed in $\mathbf{k \cdot p}$ calculation, following Eqs.~\ref{eq:layerham}--\ref{eq:soc}.
    These parameters are obtained by fitting the polarization from the low-energy model to Fig.~\ref{fig:polarization}\panela, and hence are different from those in Ref.~\citenum{du2018band}.}
    \label{tab:kdotp}
\end{table*}

\end{document}